\documentclass[12pt,preprint]{aastex}
\shorttitle{Synchrotron aging in SN 1993J}
\shortauthors{Chandra et al.}

\begin{document}

\title{Synchrotron aging and the radio spectrum of SN 1993J}

\author{P. Chandra}
\affil{Tata Institute of Fundamental Research,
Mumbai 400 005, India; {\tt poonam@tifr.res.in}}
\affil{Joint Astronomy Programme, Indian Institute of Science,
    Bangalore 560 012, India}

\author{A. Ray}
\affil{Tata Institute of Fundamental Research,
Mumbai 400 005, India; {\tt akr@tifr.res.in}}
\and
\author{S. Bhatnagar}
\affil{National Radio Astronomy Observatory, Socorro, NM; 
{\tt sbhatnag@aoc.nrao.edu}}
\affil{National Centre for Radio Astrophysics,
Pune, 411 007, India}

\begin{abstract}

We combine the GMRT low frequency radio observations of SN 1993J
with the VLA high frequency radio 
data to get  a near simultaneous spectrum around day 3200 since explosion.
The low frequency measurements of the supernova determine the turnover
frequency and flux scale of the composite spectrum and help reveal
a steepening in the spectral index, $\Delta \alpha \sim 0.6$,
in the optically thin part of the spectrum.
This is the first observational evidence of a break in 
the radio  spectrum of a young supernova.
We associate this break 
with the phenomenon of synchrotron aging of radiating electrons.
From the break in the spectrum
 we calculate the magnetic field in the shocked 
region independent of the equipartition assumption between
 energy density of relativistic particles and magnetic energy density. 
We  determine the ratio of these two energy
densities and find that this ratio is in the range:
$8\times 10^{-6}-5\times 10^{-4}$.  
We also predict  the nature of the evolution of the 
synchrotron break frequency 
with time, with competing effects due to diffusive Fermi acceleration
and adiabatic expansion of the radiative electron plasma.

\end{abstract}


\section{Introduction}

The radio spectrum of a young supernova probes the conditions in
the magnetized plasma where this radiation originates from relativistic 
electrons, which are believed to be
accelerated in the interface region of the supernova blast-wave
shock and the circumstellar medium.  The radio emission from the nearby
supernova SN 1993J is clearly
of a non-thermal nature and is argued to be due to synchrotron radiation in a
magnetic field amplified in the interaction region and affected by
synchrotron self absorption and external free-free absorption (see
e.g.\citet{fra98} and references therein).
The most critical parameter of the plasma which affects the synchrotron
radiation spectrum, is the strength of the magnetic field. This is often
estimated indirectly under assumptions of equipartition of energy between 
the magnetic fields and that of relativistic particles or by fitting the
radio flux density and turnover wavelength.
In many classical radio sources, such as supernova remnants (SNRs) like
the Crab or Cassiopeia A, or in luminous radio galaxies, the radio spectral
index is found to steepen at high frequencies 
(see e.g. \citet{kar62,eil97,mye85}). This is due to the 
so called synchrotron aging of the source, as during the lifetime of
the source, electrons with high enough energies in a homogeneous magnetic
field will be depleted due to efficient synchrotron radiation compared
with the ones with lower energies. 
An observation of a synchrotron break can yield a measurement 
of the magnetic field  {\it independent
of the equipartition argument} if the age of the source is known 
(magnetic field in the 
crab nebula was measured by \citet{pik56} (as quoted in \citet{shk60})
using this technique).

Multi-frequency radio studies of a supernova like SN 1993J, that is
bright enough in the radio bands can offer such a possibility. 
SN 1993J exploded on March 28, 1993. It was the archetypal type IIb 
supernova and provided a good opportunity to study the extragalactic 
supernovae in detail for being the nearest extragalactic supernova 
(3.6 Mpc).
In this paper, we discuss the near simultaneous 
spectrum of SN 1993J obtained by combining the GMRT low frequency data
with the VLA high frequency data around day 3200 since explosion.
We find a steepening of the spectrum by $\Delta \alpha=0.6$
at radio frequency of 4 GHz. We associate this break with the 
synchrotron cooling. Synchrotron aging  of young supernovae
has been discussed by \citet{fra98} and also mentioned by \citet{per02}. 
However we find the first clear observational signature of a 
spectral break in radio bands of the young supernova SN 1993J.
With this break frequency and the independently known age of the SN 1993J, 
we determine the magnetic field in the supernova. 
Moreover, we predict how this break frequency will
evolve with time, based on a quasi-static evolution of the 
electron spectrum under the combined effect of
acceleration processes, synchrotron losses and 
adiabatic expansion of the shell where
the electrons are confined.  We compare this observationally determined
field with the best fit magnetic field under
equipartition assumption and thence derive the fraction by 
which relativistic energy density deviates from  magnetic energy density.

We briefly describe the observations of SN1993J in section 2.
In section 3 we combine the data with the VLA data and explore
spectral fits with and without synchrotron cooling breaks.
In section 4 we explore the cumulative effects of 
adiabatic expansion of the
supernova envelope and energy gain undergone by electrons under
diffusive particle acceleration upon the 
synchrotron cooling affected particle spectrum.
In section 5 we discuss the
evolution of the break frequency with time and the importance of wide-band
spectra for modeling.

\section{GMRT Observations}

 We observed SN 1993J with the Giant Meter-wave
Radio Telescope 
in 610 and 235 MHz wavebands on 2001 December 30
and in 1420 MHz band on 2001 October 15.
 We combined this
dataset with the high frequency VLA observations (Kindly 
provided by \citet{sto03})
 in 22.5, 14.9, 8.4, 4.8 
and 1.4 GHz wavebands observed on 2002 January 13 (See Table \ref{tab:1} 
for details). 
Since the supernova is 10 years old, we do not expect the flux density to
change by significant amount in time scale of couple of months. 
For example using
time dependence of the flux density as $F\propto t^{-0.898}$ \citep{van94}, 
we find that the flux density of supernova will
change by only 2\% from 2001 October 15 to 2001 December 30, i.e. 
the flux density of SN at 1.4 GHz which was 33.9 mJy on 2001 October 15
scaled to 33.2 mJy on 2001 December 30. The difference is well within
the errors quoted.
Flux calibrators
3C48 and 3C147 were observed 
at the beginning and at the end of the observations. 
In 1420 MHz observations, we used 1034+565 as a phase calibrator, whereas,
0834+555 was used as phase calibrator for 610 and 235 MHz band observations. 
Phase calibrator was observed for 6 minutes after every 25 
minutes observation on the SN. 
The total time spend on supernova varied from 2 hours to 4 hours.
The data was analyzed using Astronomical Image Processing Software (AIPS).
The flux density errors quoted in our GMRT datasets  are
$$ \sqrt{[( {\rm map\, rms})^2 + (10\%\, 
{\rm of\,the\,peak\,flux\,of\,SN})^2]}$$
The 10\% of the peak flux takes into account any possible systematic errors 
in GMRT as well as the variation in flux in the short time 
difference in the near simultaneous spectrum.
More details of observations, data analysis and low frequency spectra
of SN 1993J at other epochs will be described elsewhere \citep{cha03}.

\section{Modeling the composite radio spectrum}

\begin{figure}
\plotone{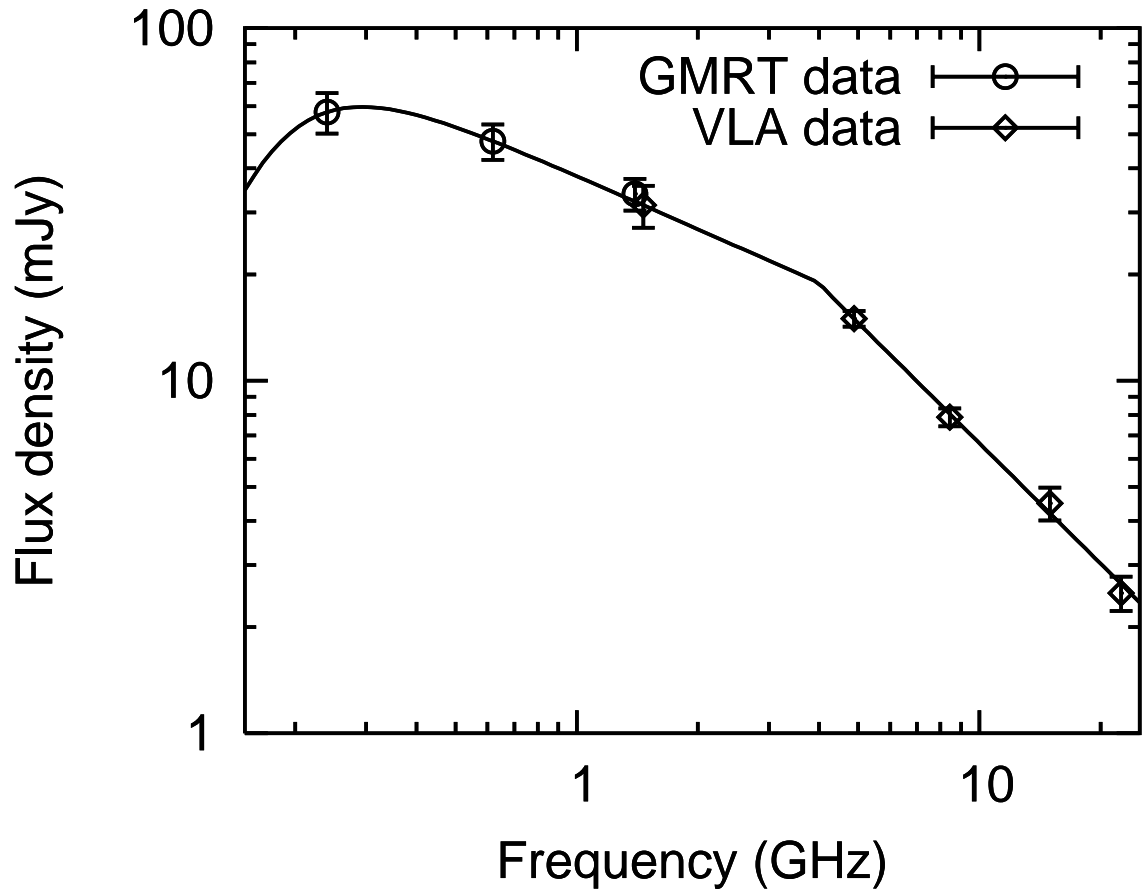}
\caption{ The spectrum of SN 1993J with synchrotron self-absorption
with a break in the spectral index at high frequencies. 
At low frequencies, the spectral emission index $\alpha$ is 0.51 
before break and after the
break, in high frequency regime, it is 1.13. \label{fig1}}
\end{figure}

When we try to fit the combined GMRT plus VLA spectrum using synchrotron 
self absorption (SSA) \citep{che98} or free free absorption (FFA) 
\citep{van94} models, we find that
the optically thin part of the spectrum can not be fitted with
a single power law. The data suggests a steepening of
spectrum towards higher frequency end and two power laws are required to
fit the optically thin part of the spectrum.
The normalized $\chi^2= 7.3 $ per $\,5 d.o.f.$ for 
SSA with a single power-law improves
significantly to $\chi^2=0.1$ per $\,3 d.o.f.$  
with SSA with  power law with a break.
We use SSA model to derive the magnetic field 
from the turn-over in the spectrum
\footnote{Based on the observational data for various radio supernovae, 
\citet{sly90} had argued that for all considered supernovae SSA is responsible 
 for the turn-over in the spectrum, even though other absorption
processes are at work.}. 
The best fit magnetic field at the peak of the spectrum
under equipartition assumption is $38.3\pm 17.1$ mG and
radius is $(1.81 \pm 0.07) \times 10^{17}$ cm. The break in the
spectrum occurs at $4.02\pm 0.19$ GHz.
Fig. \ref{fig1} shows the spectrum with the 
SSA fit with a break in the spectrum. 
We see that the emission spectral index $\alpha$ for frequency 
region below 4.02 GHz is
$\alpha=0.51\pm0.01$ (or $ \gamma=2.01\pm 0.02$, where $\gamma$ is the
particle spectral index related to $\alpha$ as, $\gamma=2 \alpha+1$). 
The   spectral index at 4.02 GHz towards higher frequency
end changes by $\Delta \alpha = 
 0.62 \pm 0.04$, hence emission spectral index after break towards
higher frequency region is $\alpha =1.13\pm 0.05$. 
%(i.e. corresponding 
%$\gamma = 2 \alpha= 2.26 \pm 0.10$). 
This variation in spectral index is roughly
consistent with that
predicted from the synchrotron cooling effect 
with continuous injection (see \citet{kar62}) and we attribute the
break in the spectrum of SN 1993J to synchrotron cooling.

\section{Synchrotron aging and effect of other energy loss/gain processes}

The lifetime of the relativistic electrons undergoing synchrotron losses is 
given as
\begin{equation}
\tau = E/[-{(dE/dt)}_{Sync}]=1.43 \times 10^{12} B^{-3/2} {\nu}^{-1/2}\; {\rm sec}
\end{equation} 
Here we use  ${B_{\perp}}^2={(B\, {\rm Sin \theta})}^2 =(2/3)B^2$.
The above expression 
is implicitly a function of time, since the magnetic field in the
region of emission itself 
changes with time as the supernova shock moves out farther
into the circumstellar plasma.
The time variation of the synchrotron break frequency can be obtained
by setting: $\tau = t$, whence,

\begin{equation} \label{sync}
%{{\nu}_{break}} = {\left(\frac{t}{1.43 \times 10^{12}}\right)}^{-2} B^{-3}
{{\nu}_{break}} = {(t/1.43 \times 10^{12})}^{-2} B^{-3}
=2 \times 10^{24} B_0^{-3}\, t
\; {\rm Hz}
\end{equation}

Here we use $B=B_0/t$ \citep{fra98}. 
From the above equation 
(and using ${\nu}_{break}=5.12\times 10^{18} B
 E_{break}^2 \;{\rm Hz}$ \citep{pac69})
and with break frequency 4 GHz, we get magnetic 
field $B=0.19$ Gauss for $t=3200$ days. 
However this estimate of the  magnetic field 
does not account for 
other processes like diffusive Fermi acceleration
and adiabatic losses, likely to be important for a young
supernova and affecting the break frequency.
Diffusive Fermi acceleration
and adiabatic expansion processes
do not result in change of slope of the energy spectrum 
from what it was in case of pure synchrotron losses,  but
the frequency where the synchrotron cooling break occurs 
gets shifted depending upon the strength of these two 
competing processes \citep{kar62}.
Adiabatic losses shifts
the 'break' frequency towards lower
frequency with time, whereas acceleration
processes shift the cooling break to higher frequencies.
We derive below the magnetic field under cumulative 
effect of all these processes.
Since supernova is young and expanding rapidly, the adiabatic losses 
will be given by
${dE/dt}_{Adia}=-(V/R)E=-E/t$. Here
$V$ is the expansion velocity, i.e. the 
ejecta velocity and $R$ is the radius of
the spherical shell.

In supernovae, diffusive mechanism is assumed to be the main acceleration
mechanism \citep{fra98,bal92}. In this process 
electrons gain energy every time they cross the shock front either from
upstream to downstream or vice versa.
The average fractional momentum gain per shock crossing or recrossing is:
$\Delta  =(4(v_1-v_2)/3v) $ and the average time taken to
perform one such cycle is \citep{bal92,dru83},
\begin{equation}
t_c=\frac{4{\kappa}_{\perp}}{v}\left(\frac{1}{v_1}+\frac{1}{v_2}\right) 
\end{equation}
Here $v$
is the test particle velocity, $v_1$ is the upstream velocity
and $v_2$ is the downstream velocity, and 
${\kappa}_{\perp}$ is the spatial diffusion coefficient
of the test particles across the ambient magnetic field, when
the shock front is quasi-perpendicular to the field.
In the rest frame of shock front,
$v_1=V$ and $v_2=v_1/4=V/4$ (for compression factor of 4). 
Hence the rate of energy gain  will be 
\begin{equation} \label{fermi}
{\left(\frac{dE}{dt}\right)}_{Fermi}=\frac{\Delta E}{t_c}=
\frac{E V^2}{ 20 {\kappa}_{\perp}}=\frac{E (R/t)^2}{ 20 {\kappa}_{\perp}}
\end{equation}

The break in the spectrum will occur for those
electron energies for which the time scales for the  
cumulative rate of change of electron energy
due to synchrotron cooling plus adiabatic losses
and gain through diffusive acceleration 
becomes comparable to the life time of the supernova \citep{kar62}.
Lifetime of electrons  for the cumulative energy loss rate is
\begin{equation}
\tau = \frac{E}{{(dE/dt)}_{Total}}=\frac{E}{ (R^2 t^{-2}/20 {\kappa}_{\perp})E
-b B^2  E^2- t^{-1}E}
\end{equation}
where first term in the denominator is synchrotron loss term with 
$b=1.58 \times 10^{-3}$. Setting
the life time $\tau=t$, break frequency can be derived as: 
\begin{equation} 
\label{synchr_freq} 
{\nu}_{break}=\frac{2\times 10^{24}}{{B_0}^3}
{\left[\frac{R^2}{20 {\kappa}_{\perp}}t^{-1/2}-2\,t^{1/2}\right]}^2
 \,{\rm Hz}
\end{equation}

We do not yet have an observational determination
of ${\kappa}_{\perp}$ for SN 1993J.
However in the case of SN 1987A, from the delay in the
switch-on of the emission between
843 MHz and 4.8 GHz, this parameter was estimated to be
${\kappa}_{\perp}=2\times 10^{24}$ cm$^2$s$^{-1}$
\citep{bal92}. This  
is relatively independent of the density and magnetic field of the two
clumps in the CSM into which the SN 1987A shock was running
into in the first 1500-1600 days. 
We use a slightly different $\kappa_{\perp}$ for SN 1993J, since
there is evidence that the compression ratio $\rho$ of gas or plasma across
the shock in SN 1993J is higher than that of SN 1987A
\footnote{Diffusive acceleration predicts a flattening of the 
spectrum at a spectral index $\alpha = 3/[2(\rho -1)]$; 
which gives 
$\rho =4$ in case of SN 1993J in contrast to $\rho = 2.7$ for SN 1987A 
\citep{bal92}.}.
Therefore, we estimate that ${\kappa}_{\perp}$ for SN 1993J is
 (scaled by the higher compression ratio in SN 1993J):
${\kappa}_{\perp}=\frac{4}{2.7}\times
2 \times 10^{24}$ cm$^2$s$^{-1}$ i.e. $2.96 \times 10^{24}$ cm$^2$s$^{-1}$.
From VLBI observations \citep{bar02}, 
the extrapolated angular radius of SN 1993J
on day 3200 is $\sim 5012\, {\rm \mu as}$ i.e. $\sim 2.65\times 10^{17}$ cm. 
Using ${\kappa}_{\perp}$ and R in Eq. \ref{synchr_freq} we find magnetic field
$B=0.33\pm0.01 $ Gauss, from the observationally determined break.

From our SSA best fit to the radio spectrum, 
we have a value of the best fit 
equipartition magnetic field $B_{eq} = 38.2\pm17.1$ mG.
If $a$ is the ratio between relativistic electron energy density, 
$ U_{rel}$,
to magnetic energy density, $ U_{mag}$, ($a=U_{rel}/U_{mag}$), 
then magnetic field $B$ depends on equipartition
factor $a$ as $B \propto {a}^{-4/(2 \gamma +13)}$ (see \citet{che98}).
Hence, we get equipartition fraction $a$ i.e.  
$U_{rel}/U_{mag} =  8.5 \times 10^{-6}-5.0 \times10^{-4}$. 
This small value
of the ratio of particle energy density vs magnetic energy density indicates
that the plasma kinetics is dominated by magnetic field and associated
turbulence.
                                                                                
\section{Discussion and Conclusions}

We see in the last section that the magnetic field
calculated from the break in the spectral index
is $0.33 $ Gauss, which is
$\sim$1.4 times higher than that expected from an extrapolation of 
\citet{fra98} ($B=0.24$ Gauss at day 3200). If we took only the
synchrotron cooling effect and neglected adiabatic expansion and
diffusive Fermi acceleration, we get magnetic field $B=0.19$ Gauss,
which is in closer agreement with that of \citet{fra98}.
However, at this young age of the supernova the effects of
adiabatic expansion and diffusive Fermi acceleration are likely
to be significant as seen in SN1987A \citep{bal92} and
hence these effects should not be neglected.

One can estimate the importance of the diffusive acceleration term, if 
one is able to follow how the break frequency evolves with time.
In Eq. \ref{synchr_freq}, first term is the contribution of the acceleration
and second term is the contribution of adiabatic expansion and synchrotron
losses.  We note that for the estimated value of 
${\kappa}_{\perp}$ at the present epoch, diffusive Fermi acceleration 
dominates and will continue to dominate over the 
adiabatic losses until about 20 years since explosion.
Therefore, at present epoch the break frequency evolves as 
${\nu}_{break} \propto t^{-1}$. 
After 20 years when the acceleration will cease 
to dominate over adiabatic expansion and  
the break frequency will increase as: 
${\nu}_{break} \propto t$. 

Since we do not have an independent method to 
estimate the spatial diffusion coefficient 
${\kappa}_{\perp}$ for SN 1993J, we have used the linearly scaled
 (by the respective  compression 
ratios, $\rho$) value of ${\kappa}_{\perp}$ measured for
SN 1987A from direct radio observations (see \citet{bal92}). 
However, we can directly calculate the 
value of  ${\kappa}_{\perp}$ from the (measurable) rate
of change of ${\nu}_{break}$, i.e. from the expression:
\begin{equation}
\frac{d {\nu}_{break}}{dt} = \frac{2 \times 10^{24}}{ B_0^{3}} 
\left[\frac{R^2 }
{20 {\kappa}_{\perp}}t^{-1/2}\,-2\,t^{1/2}\right]\left[-\frac{R^2 }
{20 {\kappa}_{\perp}}t^{-3/2}\,-2\,t^{-1/2}\right] {\rm Hz/sec}
\end{equation}
For the present epoch and with the estimated parameters as above, we calculate 
that the break frequency is changing at the rate of 1.2 GHz/year.
Thus a few more multi-frequency spectral observations across GMRT and
VLA bands, 
separated by a few years, will observationally determine the temporal
variation in the break frequency 
and underline the importance of the diffusive acceleration effects.

\begin{figure}
\plotone{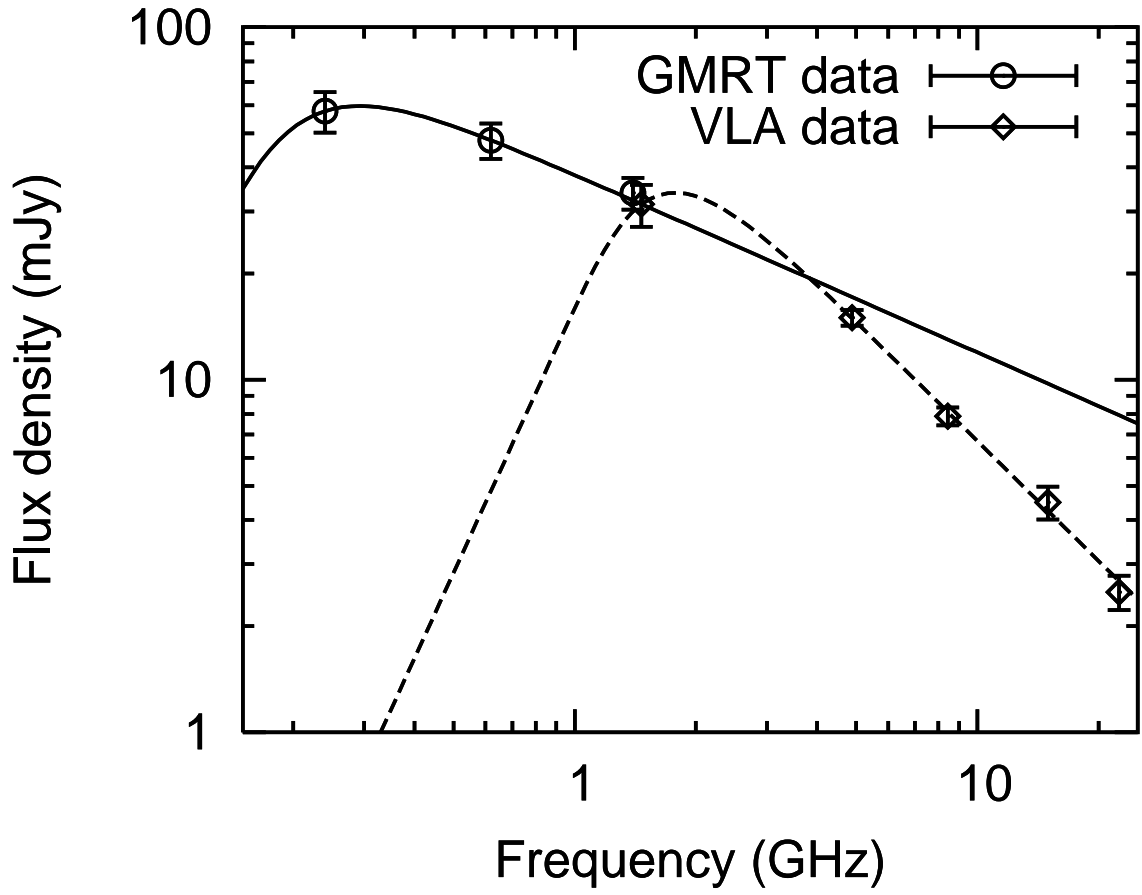}
%\plottwo{f2a.eps}{f2b.eps}
\caption{Wide band spectrum of SN 1993J. Synchrotron self absorption 
fit to "only" low frequency data
(solid line) and "only" high frequency data (dashed line).\label{fig2}}
\end{figure}

The combination of multi-frequency
radio spectrum across GMRT and VLA bands
is critical for deriving the above results.
In Fig. \ref{fig2} we show a comparison of synchrotron self absorption
model (with a single optically-thin power-law index) fitted 
only to the low frequency data (0.22 GHz to 1.4 GHz) versus such a model
fit obtained with only the higher frequency data (1.4 GHz to 22.5 GHz).
This comparison shows that while the model fitted only to 
the low frequencies over-predicts
the flux density at high frequencies, the model
fitted only to high frequencies on the other hand fails 
to account for both synchrotron cooling break and 
 seriously under-predicts the low frequency flux densities.
The comparison underscores the importance of broad band observations
for determining the physical processes taking place in the supernova.

We note the uncertainity in 
%the spatial diffusion coefficient 
${\kappa}_{\perp}$
for SN 1993J as also emphasised by the referee who suggests that
${\kappa}_{\perp}$ may not be a constant, and points out a paper 
due to \citet{rey98}. This paper {\it assumes} that 
${\kappa}_{\perp}$ is proportional to the particle energy. This dependence may
affect the determination of the magnetic
field (see Eq. \ref{synchr_freq}). Only future observations of the
break frequency evolution
with time will directly put constraints on ${\kappa}_{\perp}$ for SN 1993J.
We also note that our results are based on the assumption
that the acceleration and synchrotron losses are taking place in
the same region. However, if the regions of the two processes are not
substantially overlapping, the synchrotron break frequency will not be affected
by acceleration. Even in that case, the magnetic field is
much higher than the equipartiton
magnetic field and the plasma is still dominated by the magnetic
energy density.

\acknowledgements 

We thank Kurt Weiler and Christopher Stockdale
for providing us the high frequency VLA data. 
We thank the NRAO staff for providing AIPS.
We thank the staff of the GMRT that made these observations possible. 
GMRT is run by the National Centre for Radio Astrophysics of the 
Tata Institute of Fundamental Research. We thank the anonymous referee
for useful comments.
% All of this work was done using 
%computers running the GNU/Linux operating system and we thank all the 
%numerous contributors to this software.

\clearpage

\begin{table}
%\scriptsize
\caption{Observations of the spectrum of SN 1993J
\label{tab:1}}
\begin{tabular}{ccccc}
\tableline\tableline
Date of  & Days since & Frequency &  Flux density & rms\\
Observation&  explosion & in GHz  & mJy   & mJy \\
\tableline\tableline
Dec 31,01                  & 3199 & 0.239  & 57.8 $\pm$7.6   & 2.5\\
Dec 30,01                  & 3198 & 0.619  & 47.8 $\pm$5.5   & 1.9\\
Oct 15,01                  & 3123 & 1.396  & 33.9 $\pm$3.5   & 0.3\\
Jan 13,02\tablenotemark{a} & 3212 & 1.465  & 31.44$\pm$4.28  & 2.9\\
Jan 13,02\tablenotemark{a} & 3212 & 4.885  & 15   $\pm$0.77  & 0.19\\
Jan 13,02\tablenotemark{a} & 3212 & 8.44   & 7.88 $\pm$0.46  & 0.24\\
Jan 13,02\tablenotemark{a} & 3212 & 14.965 & 4.49 $\pm$0.48  & 0.34\\
Jan 13,02\tablenotemark{a} & 3212 & 22.485 & 2.50 $\pm$0.28  & 0.13\\
\tableline\tableline
\end{tabular}
\tablenotetext{a} {VLA data points, \citep{sto03}}
\end{table}

\clearpage

\clearpage                                                                                

\end{document}